\documentclass[prb,preprint]{revtex4}

\usepackage{graphicx}
\usepackage{amsmath}

\begin{document}

\title{Figure of Merit of one-dimensional resonant transmission systems in the quantum regime}

\author{Mustafa Ali \c{C}ipilo\u{g}lu\footnote{Corresponding
     author: e-mail: {\sf ali.cipiloglu@bayar.edu.tr}, }}
\address{Department of Physics, Celal Bayar University,\\
 45037 Manisa, Turkey}

\author{Sadi Turgut\footnote{Corresponding
     author: e-mail: {\sf sturgut@metu.edu.tr}}}
\address{Department of Physics, Middle East Technical University,\\
06531, Ankara, Turkey}

\begin{abstract}
The figure of merit, $ZT$, for a one-dimensional conductor
displaying a Lorentzian resonant transmission probability is
calculated. The optimum working conditions for largest $ZT$ values
are determined. It is found that, the resonance energy has to be
adjusted to be several resonance widths away from the Fermi level.
Similarly it is better for the temperature to be equal to several
resonance widths. The approximate relationships, which can be a
fairly good guide for designing devices, between different
parameters under optimum working conditions are given.
\end{abstract}

\maketitle

\section{Introduction}

Recent advances in fabrication and material growth technologies have
made possible the production of devices whose dimensions are order
of a few nanometers. The thermoelectric power, thermal and
electrical conductivities of these mesoscopic scale materials are of
interest for thermoelectric device applications\cite{harman}, such
as heat pumps\cite{hongkai} and power generators. The performance of
a thermoelectric device is usually quantified by a dimensionless
number called as figure of merit, $ZT$, which measures the
efficiency of thermoelectric energy conversion. Higher values of
$ZT$ correspond to higher thermoelectric energy conversion
efficiency so that, for example, if $ZT$ tends to infinity, the
efficiency approaches to that of an ideal Carnot engine. Recent work
on superlattice semiconducting devices demonstrated $ZT \approx 2.5$
at room temperature\cite{venkat} and $ZT \approx 1.4$ at high
temperatures\cite{saramat}, breaking the long-standing limit of $ZT
\approx 1$ for most of best known thermoelectric materials.

In 1D nanoscale systems, the increase of $ZT$ might be further
enhanced by the quantum confining of electrons and phonons in low
dimensions\cite{hicks}. In the recent experimental study, for
instance, $ZT$ could be increased by embedding nanoparticles in a
crystalline semiconductors\cite{woochul}. High values of $ZT$ can
also be obtained for one-dimensional structures displaying resonant
transmission. In this case, the transmission probabilities are very
sensitive to the electron energies leading to large values of the
thermopower and hence of the figure of merit. In this contribution,
relationships between different thermoelectric coefficients are
calculated, and probably the first, $ZT$ values of a resonant
tunnelling device is computed and its dependence on device
parameters is investigated. We are expecting that our calculations
will be a good guide to researchers studying on thermoelectric
devices.

\section{Transport coefficients}
In here a one-dimensional mesoscopic device is considered. It is
assumed that there is only one transverse mode that is occupied by
the electrons. The thermoelectric currents in such a device under
linear regime can be expressed in terms of the energy dependent
transmission probability $\mathcal{T}(E)$ of the
electrons\cite{sivanimry}. The electric current, $I$, and the heat
current, $\dot{Q}$, under a potential difference $\Delta V$ and a
temperature difference $\Delta T$ can be expressed as\cite{cipil}
\begin{eqnarray}
I &=& \frac{2e^2}{h}g_0 \Delta V + \frac{2(-e)k_B}{h}g_1 \Delta T \quad, \\
\dot{Q} &=& \frac{2(-e)k_BT}{h}g_1 \Delta V + \frac{2k_B^2T}{h}g_2
\Delta T \quad,
\end{eqnarray}
where $g_n$ are
\begin{equation}
g_n= \int_{-\mu/k_BT}^{\infty} dx~ x^n \left( -f^\prime(x)\right)
  \mathcal{T}(\mu+xk_BT) \quad,
\label{gn}
\end{equation}
$\mu$ is the chemical potential and $f(x)=1/(1+e^x)$ is the
Fermi-Dirac distribution function.

Frequently measured transport coefficients, the electrical
conductance $G_{\textrm{el}}$, the thermal conductance
$G_{\textrm{th}}$, the Seebeck coefficient $S$, and the
dimensionless thermoelectric figure of merit $ZT$ can be expressed
as
\begin{eqnarray}
G_{\textrm{el}} &=& \frac{2e^2}{h} g_0 \quad,\\
G_{\textrm{th}} &=& \frac{2k_B^2}{h} T\left( g_2-\frac{g_1^2}{g_0}
    \right)\quad,\\
S &=& \frac{k_B}{(-e)}\frac{g_1}{g_0} \quad,\\
ZT &=& \frac{S^2 G_{\textrm{el}}T}{G_{\textrm{th}}}=
\frac{g_1^2}{g_0 g_2 - g_1^2}\quad.
\end{eqnarray}

It can be seen that, to obtain large values of $ZT$, the factor
$g_1$, which also appears in the Seebeck coefficient, has to be
large. Equation~(\ref{gn}) implies that at low temperatures, $g_1$
is basically proportional to the first derivative of
$\mathcal{T}(E)$. For this reason, large values of $g_1$ can be
achieved when $\mathcal{T}(E)$ is strongly energy dependent, and the
fastest change in it occurs around the Fermi level. In this
contribution we investigate a resonant tunnelling device where the
transmission probability is assumed to have a Lorentzian form
\begin{equation}
\mathcal{T}(E)=\frac{\mathcal{T}_{\textrm{max}}}{1+(E-E_o)^2/\Gamma^2}\quad,
\label{T_of_E}
\end{equation}
where $E_o$ is the resonance energy and $\Gamma$ is the half width.
The factor $g_1$ and the Seebeck coefficient $S$ changes sign when
the Fermi energy crosses the resonance energy $E_o$. However, when
$\mu$ differs from $E_o$ by an energy of the order of $\Gamma$, both
of these quantities attain large values and this is the region where
we should look for large values of $ZT$. It is assumed that the
Fermi level is very large compared to the temperature so that the
lower limit of the integral in Eq.~(\ref{gn}) is extended to minus
infinity.

\section{Results and discussion}

It can be seen that the values of $ZT$ depend on two dimensionless
parameters. One of them, $\epsilon=(\mu-E_o)/\Gamma$, indicates the
distance of the Fermi level from the resonance energy in units of
the half width $\Gamma$, and the other, $\theta=k_BT/\Gamma$ gives
the temperature compared to the resonance width.

\begin{figure}
\includegraphics[scale=0.80]{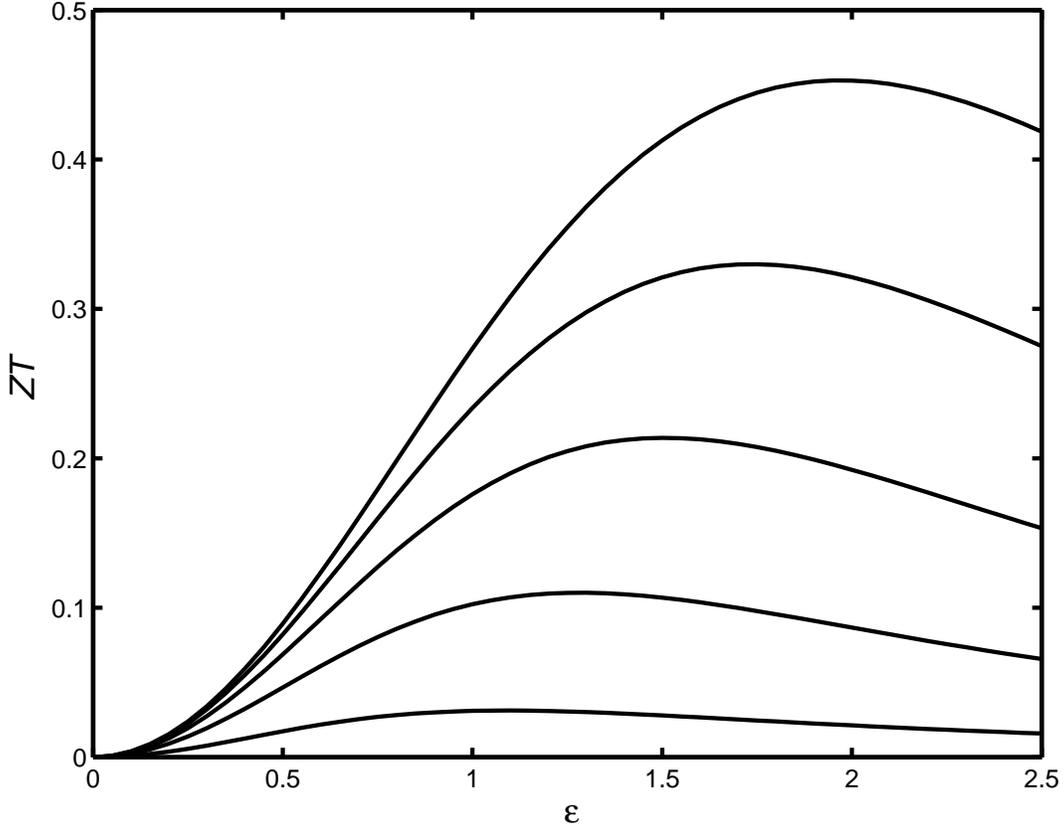}
\caption{ \label{fig1} Variation of $ZT$ for different values of
temperature ($\theta=$0.1, 0.2, 0.3, 0.4, 0.5 from bottom to top) as
a function of $\epsilon$. }
\end{figure}

The values of $ZT$ as a function of dimensionless Fermi level,
$\epsilon$, for different temperatures is shown in Fig.~\ref{fig1}.
The Seebeck coefficient is zero at $\epsilon=0$ and has different
signs at different sides of this point. Since $ZT$ is an even
function of $\epsilon$, only the positive values of $\epsilon$ is
shown in the figure. It can be seen that when $\epsilon$ is varied,
$ZT$ values increase, reach a maximum and then start to decrease.
The place and the value of the maxima depends on the temperature.

\begin{figure}
\includegraphics[scale=0.45]{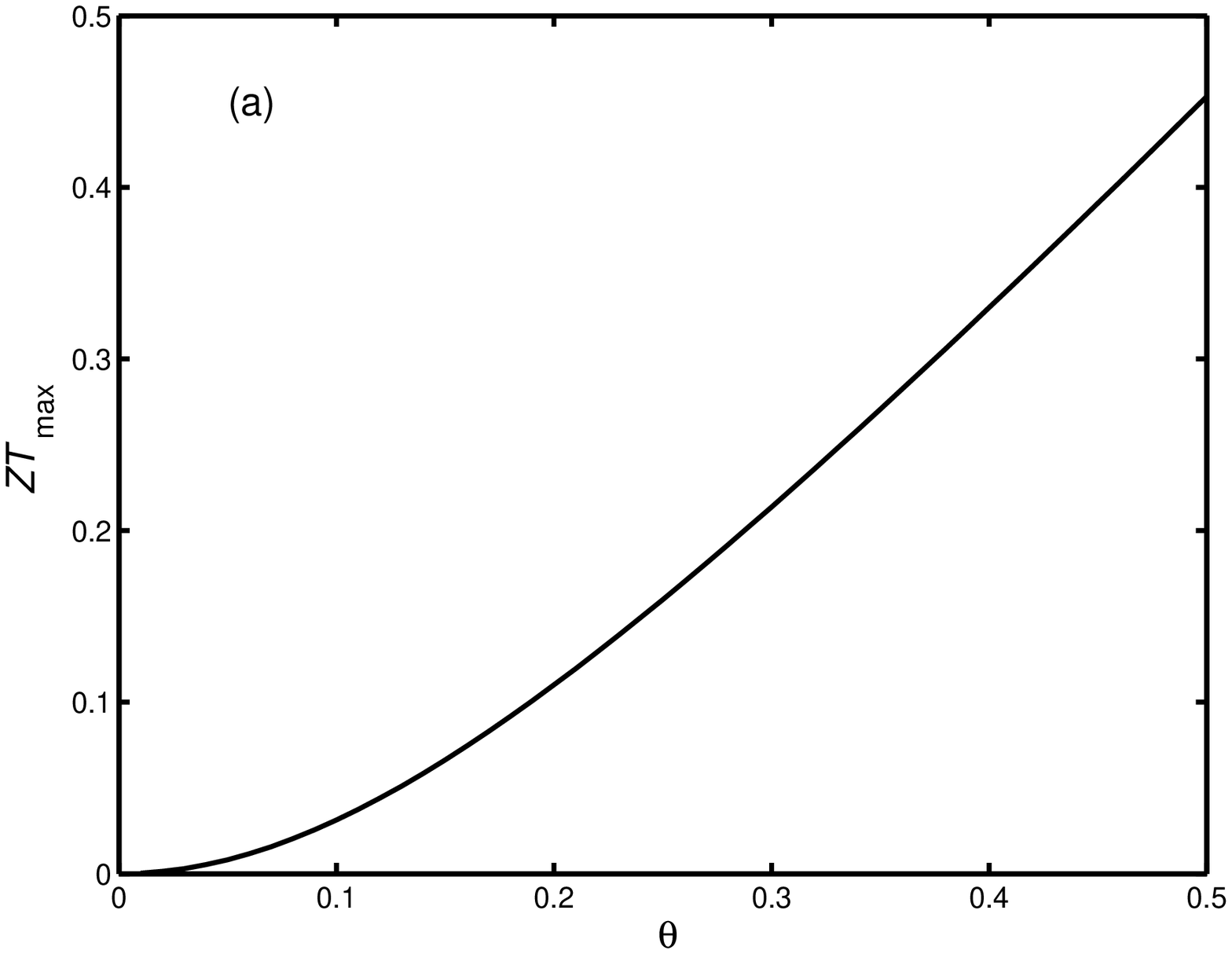}
\includegraphics[scale=0.45]{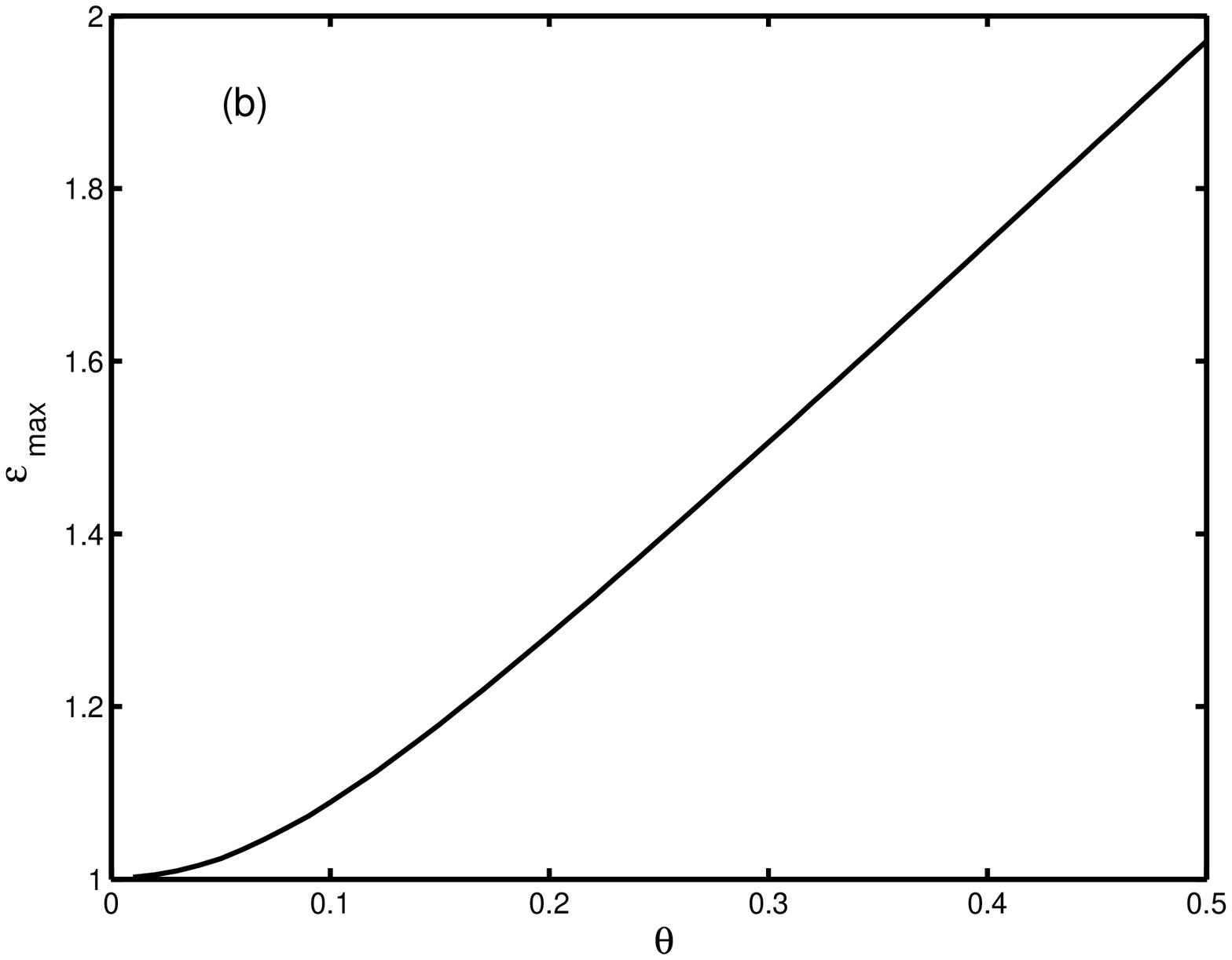}
\caption{ \label{fig2} (a) The maximum values of $ZT$ as a function
of $\mu$ is plotted for different values of $\theta$. (b) The value
of $\epsilon$ at maxima is plotted for different values of $\theta$.
}
\end{figure}

The maximum attainable value of figure of merit, $ZT_{\textrm{max}}$
and the best place of Fermi level, $\epsilon_{\textrm{max}}$ for
each temperature are plotted in Fig.~(\ref{fig2}). For low
temperatures, $\lesssim 0.1$, the best place of the Fermi level is
approximately one resonance width above or below of the resonance
energy ($\epsilon\approx\pm1$), but the maximum attainable figure of
merit value is very low. For high temperatures however, both of
these parameters have a linear dependence on temperature which are
approximately given as
\begin{eqnarray}
\epsilon_{\textrm{max}} \approx 0.8 + 2.4 \theta \quad,\\
ZT_{\textrm{max}} \approx 1.4\theta-0.3\quad.
\end{eqnarray}
It can be seen that, in each case the optimum value of the figure of
merit is obtained when $\mid\epsilon\mid>1$, i.e., the Fermi level
should be more than one half-width away from the resonance energy.
If the Lorentzian form of the resonance in Eq.~(\ref{T_of_E})
remains valid for a large energy interval, sufficiently large $ZT$
values can be obtained in this domain.

\begin{figure}
\includegraphics[scale=0.80]{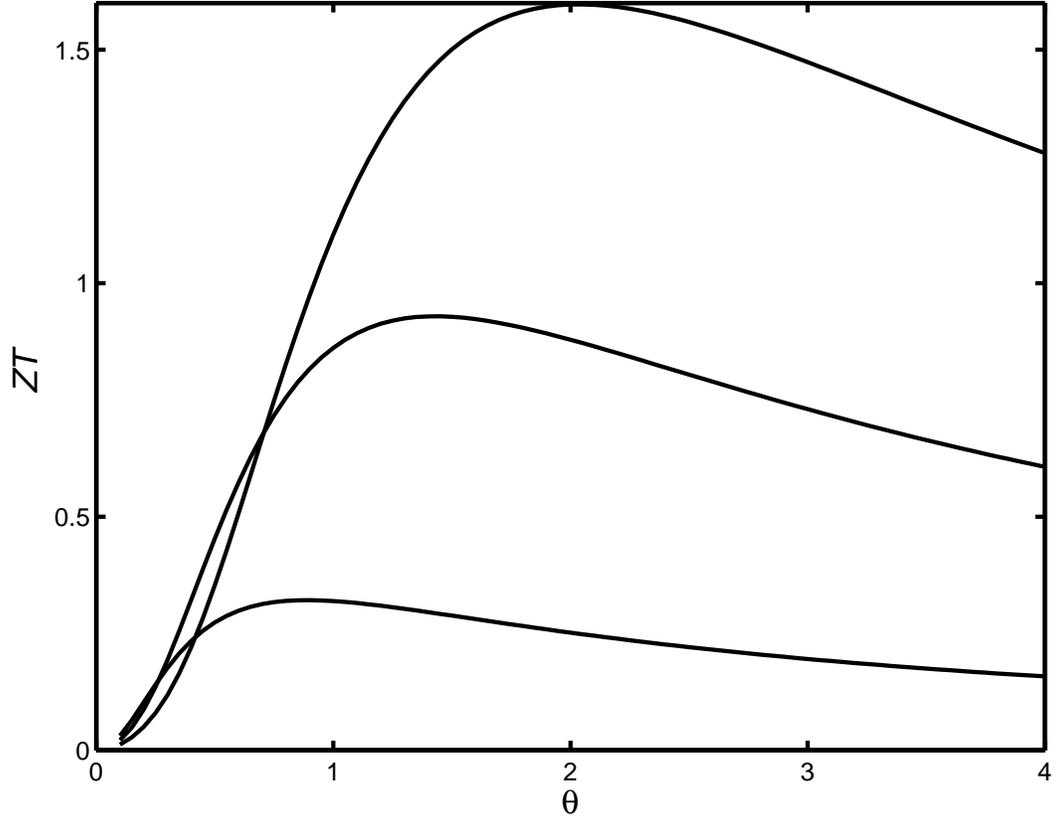}
\caption{ \label{fig3} Variation of $ZT$ for different placements of
the Fermi level ($\epsilon=$1, 2, 3 from bottom to top) as a
function of $\theta$. }
\end{figure}

\begin{figure}
\includegraphics[scale=0.45]{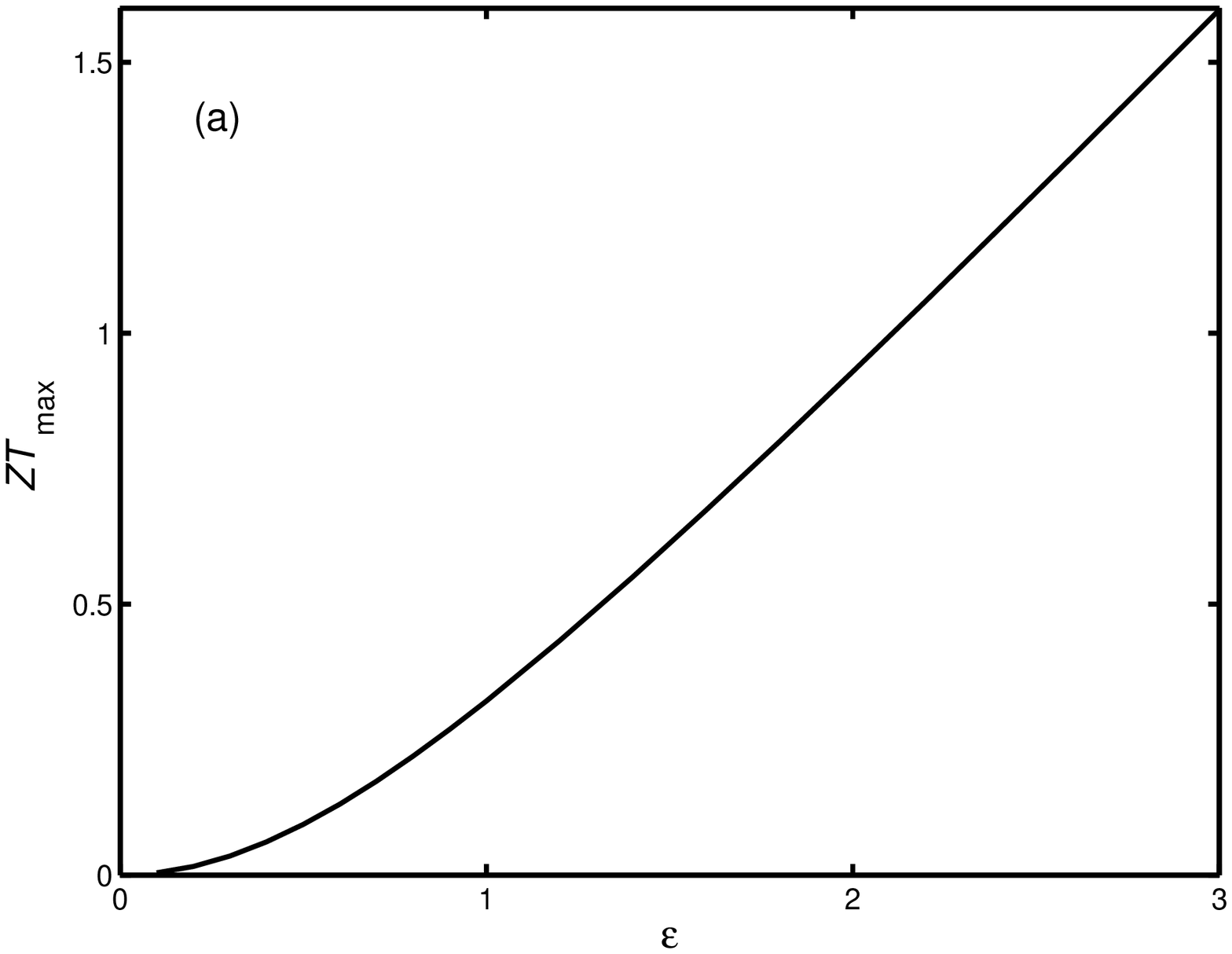}
\includegraphics[scale=0.45]{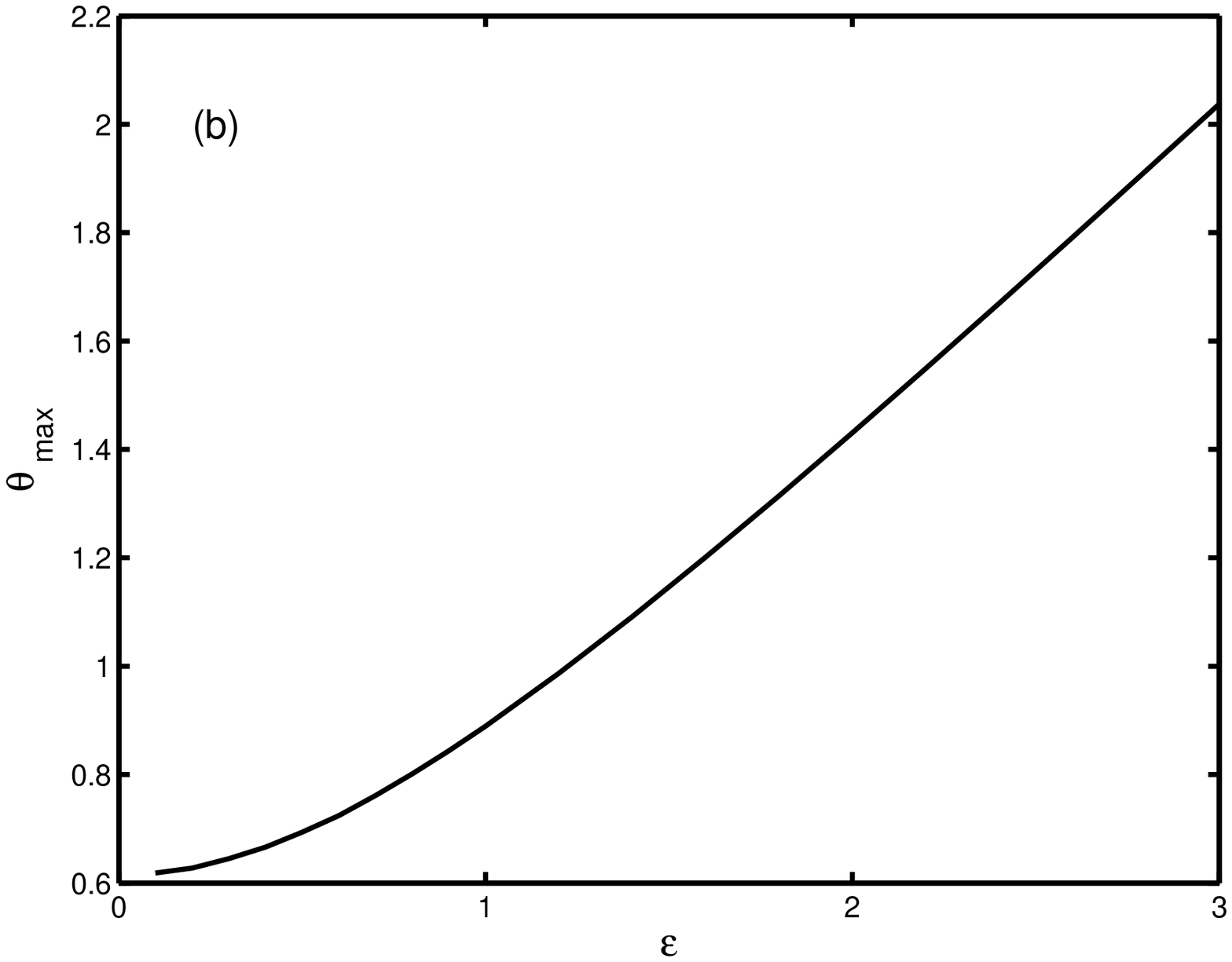}
\caption{ \label{fig4} (a) The maximum values of $ZT$ as temperature
is varied is plotted as a function of $\epsilon$. (b) The value of
$\theta$ at maxima is plotted as a function of $\epsilon$. }
\end{figure}

The values of $ZT$ as a function of $\theta$ for different places of
the Fermi level is shown in Fig.~\ref{fig3}. The figure of merit
reaches a maximum for a certain temperature in this case as well.
The maximum attainable value of figure of merit, $ZT_{\textrm{max}}$
and the optimum temperature, $\theta_{\textrm{max}}$ are shown in
Fig.~(\ref{fig4}) as a function of $\epsilon$. When the Fermi level
is almost coincident with the resonance energy ($\epsilon\lesssim
1$), the temperature at which maximum attained is around
$\theta\approx0.6$. However, in this region, $ZT$ values are low. In
the opposite case, for $\epsilon\gtrsim1$, the optimum temperature
and the best value of $ZT$ appear to be linear in $\epsilon$ with
the following numerical relationships
\begin{eqnarray}
\theta_{\textrm{max}} \approx 0.6\epsilon+0.2\quad,\\
ZT_{\textrm{max}} \approx 0.7\epsilon-0.4\quad.
\end{eqnarray}
The approximate relation $\theta_{\textrm{max}} \approx 0.9
ZT_{\textrm{max}}+0.6$ seems to hold over the whole range
investigated in this study.

It is seen that for one-dimensional systems displaying resonant
transmission, it is possible to obtain large values of the figure of
merit $ZT$ by fine-tuning the device parameters, the resonance width
and the place of the resonance energy relative to the Fermi level.
The optimum working conditions for the largest values of $ZT$
appears to be far away from the resonance: Fermi level is several
$\Gamma$ away from the resonance energy $E_o$ and $k_BT$ is several
times $\Gamma$. It can be argued that, under such extreme cases, the
validity of Lorentzian form in Eq.~(\ref{T_of_E}) is questionable
and some of the results obtained in here can be changed. However,
the crucial feature leading to these results is the strong energy
dependence of the transmission probability. Specifically, the
transmission probability should continue decreasing at the Fermi
level and not reach to a minimum. For this reason, similar results
can be expected for realistic devices. Numerical computations for a
double barrier system displaying resonances indicate that when the
half width $\Gamma$ is much smaller than the distance between the
resonances, results obtained are roughly in agreement with the ones
in here. For this reason Lorentzian form can be used as a fairly
good guide for designing devices.

An important problem, connected with being far away from the
resonance, is the smallness of the electrical and thermal
conductances. For this reason, efficient devices can only work with
very small power. Moreover, precautions should be taken to make the
phonon contribution to the heat conductance to be very small as this
will always decrease the $ZT$ value.

\section{Conclusions}

The figure of merit has been computed for a system whose
transmission probability displays a Lorentzian peak. By adjusting
the resonance width to be small and carefully placing the Fermi
level for the working temperatures of interest, significantly high
values of $ZT$ can be obtained. Approximate numerical relationships,
which can be a guide for designing thermoelectric devices, for the
optimum values of the parameters are also given.

\end{document}